\documentclass[11pt,a4paper]{article}
\pdfoutput=1

\usepackage{jheppub}

\usepackage{amsmath}
\usepackage{bbm}
\usepackage{float}
\usepackage{graphicx}	
\usepackage{hyperref}
\usepackage{mathtools}
\usepackage{cancel}

\usepackage[normalem]{ulem}

\usepackage{multirow}
\usepackage{rotating}
\usepackage{subcaption}

\def\lsim{\raise0.3ex\hbox{$\;<$\kern-0.75em\raise-1.1ex
\hbox{$\sim\;$}}}
\def\gsim{\raise0.3ex\hbox{$\;>$\kern-0.75em\raise-1.1ex
\hbox{$\sim\;$}}}

\def\thetitle{
Comparative study of the 1-2 exchange symmetries in neutrino frameworks with global and local validities  
 \vspace{- 6mm}
}
\title{\thetitle}
\hypersetup{pdftitle={\thetitle}}
\author{Hisakazu Minakata}
\affiliation{
Center for Neutrino Physics, Department of Physics, Virginia Tech, Blacksburg, Virginia 24061, USA \\
}

\emailAdd{hisakazu.minakata@gmail.com}
\date{\today}

\abstract{A new picture of ``one-resonance - one-symmetry'' has been proposed recently to reveal nature of the reparametrization symmetry in neutrino oscillation in matter and where it resides: Symmetry of $i \leftrightarrow j$ state-exchange-type exists at around a resonance, with $i$ and $j$ being the states which participate in the level crossing. Consistently, the 1-2 and 1-3 state exchange symmetries are identified at around the solar and atmospheric resonances, respectively, in the locally-valid frameworks. On the other hand, the Denton {\it et al.} (DMP) perturbation theory, a globally-valid framework, has the 1-2 exchange symmetry which is akin to the one in the aforementioned solar-resonance perturbation (SRP) theory. In our picture, the symmetry must be associated with the resonance, not the framework, and if so, these two 1-2 symmetries must be identical to each other. We conduct a comparative study of the 1-2 symmetries possessed by SRP and DMP to confirm their identity. An almost identity is verified, but in a highly nontrivial way. 
}

\begin{document} 

\maketitle

\section{Introduction}
\label{sec:introduction}

Recently, we have started investigation of a new type of symmetry in neutrino oscillation in matter which may be called as the reparametrization symmetry~\cite{Minakata:2022yvs,Minakata:2021dqh,Minakata:2021goi,Minakata:2022zua}. Invariance of the oscillation probability under such symmetry transformations implies that there is another way of parametrizing the equivalent solution of the theory. To search for such symmetry in a systematic way, we have introduced a method called {\em ``Symmetry Finder''} (SF). It has been successfully applied to the several perturbative frameworks of neutrino oscillation in matter, as summarized in ref.~\cite{Minakata:2022yvs}. 

To the author's knowledge, the first discussion of the reparametrization symmetry in neutrino oscillation is given by Fogli {\it et al.} in ref.~\cite{Fogli:1996nn}. Sometime later, there appeared a much-discussed ``light-side--dark-side'' symmetry~\cite{deGouvea:2000pqg,Fogli:2001wi}, the one between the first and the second octants of $\theta_{12}$, 
\begin{eqnarray} 
&&
m^2_{1} \leftrightarrow m^2_{2}, 
\hspace{10mm}
\theta_{12} \leftrightarrow \frac{\pi}{2} - \theta_{12}, 
\label{light-dark}  
\end{eqnarray} 
by which a growing popularity of such symmetry resulted. This ``light-dark'' symmetry is approximately realized as a result of the reactor measurement of $\theta_{12}$~\cite{KamLAND:2013rgu}. But, of course, the second octant solution is excluded if the data of the solar neutrino measurements are added. See e.g.~ref.~\cite{Super-Kamiokande:2016yck}. It is because the matter effect~\cite{Wolfenstein:1977ue,Mikheyev:1985zog} breaks the first octant - second octant degeneracy. It teaches us a lesson that we have to take the matter effect into account whenever it is relevant because it affects the symmetry. 
For our frequent usage, we denote the reparametrization symmetry of the type discussed here and in these references as the ``Rep symmetry'' for short throughout this paper. 

Despite successes of the SF method in digging out the Rep symmetries, as reported in refs.~\cite{Minakata:2022yvs,Minakata:2021dqh,Minakata:2021goi,Minakata:2022zua}, there remain many questions to be answered. One of the most important is the nature of the Rep symmetry. That is, whether it reflects physical feature of the dynamics of neutrino evolution in matter, or it is merely a framework-dependent regularity, in each particular one used to describe neutrino evolution. 

In this paper, to shed light on this issue, we engage a comparative study of the two Rep symmetries of the 1-2 state exchange type~\cite{Minakata:2022yvs,Minakata:2021dqh}, one in the solar-resonance perturbation (SRP) theory~\cite{Martinez-Soler:2019nhb} and the other in the Denton {\it et al.} (DMP) perturbation theory~\cite{Denton:2016wmg}. We ask the question, ``how and why are they so  identical to each other?'', answer of which must shed light on nature of the Rep symmetry. Even more generally outside the present context, their identity is a nontrivial issue because SRP and DMP are the very different frameworks in nature, such as the structure of perturbative expansions. For further discussions of these points, see sections~\ref{sec:SF} and ~\ref{sec:new-picture}. 

Looking back, symmetry in neutrino oscillation has been discussed in the various related or different contexts. The above mentioned symmetries under discrete transformations may have affinity with the discrete symmetries in flavor physics models~\cite{Altarelli:2010gt}. The parameter degeneracy, the problem of multiple solutions for a given set of observables, has been discussed from the viewpoint of approximate or exact symmetries of the oscillation probability~\cite{Fogli:1996pv,Burguet-Castell:2001ppm,Minakata:2001qm,Minakata:2010zn} in the $\nu$SM, a shorthand notation for neutrino-mass-embedded Standard Model, and in extensions with the non-standard interactions beyond the $\nu$SM~\cite{Coloma:2016gei}. For discussions on the related another aspects of symmetries see, e.g., refs.~\cite{Gluza:2001de,deGouvea:2008nm,Zhou:2016luk}. 

From section~\ref{sec:systematic-way} to section~\ref{sec:SF}, we will give a brief pedagogical review of the Rep symmetry in neutrino oscillation. We start from a derivation of the ``light-dark'' symmetry, in fact its slightly more generic version, in a systematic way. Then, we explain why the symmetry in matter is more profound than that in vacuum, and give a brief summary of the current status of the symmetries identified by using the SF method. It will be followed by presentation of our new picture of the SF symmetry and the discussion of nature of the Rep symmetry in section~\ref{sec:new-picture}. It naturally leads us to the comparative study of the 1-2 state exchange symmetries in the neutrino frameworks with the global or local validities, which will be carried out in section~\ref{sec:comparative}, the core part of this paper. We conclude in section~\ref{sec:conclusion}. 

\section{A systematic way of getting the Rep symmetry} 
\label{sec:systematic-way} 

There is a way of systematizing the finding path for the ``light-dark'' symmetry mentioned in Introduction. In vacuum, the flavor eigenstate 
$\nu \equiv \left[ \nu_{e}, \nu_{\mu}, \nu_{\tau} \right]^{T}$ is related to the mass eigenstate 
$\bar{\nu} \equiv \left[ \nu_{1}, \nu_{2}, \nu_{3} \right]^{T}$ using the flavor mixing matrix $U \equiv U_{\text{\tiny MNS}}$~\cite{Maki:1962mu} as $\nu = U \bar{\nu}$. Interestingly, the flavor-mass eigenstates relation can take the following three different forms~\cite{Parke:2018shx} 
\begin{eqnarray} 
\left[
\begin{array}{c}
\nu_{e} \\
\nu_{\mu} \\
\nu_{\tau} \\
\end{array}
\right] 
&=& 
U_{23} (\theta_{23}) U_{13} (\theta_{13}) 
U_{12} (\theta_{12}, \delta) 
\left[
\begin{array}{c}
\nu_{1} \\
\nu_{2} \\
\nu_{3} \\
\end{array}
\right] 
= 
U_{23} (\theta_{23}) U_{13} (\theta_{13}) 
U_{12} \left( \theta_{12} + \frac{\pi}{2}, \delta \right) 
\left[
\begin{array}{c}
- e^{ i \delta } \nu_{2}  \\
e^{ - i \delta } \nu_{1} \\
\nu_{3} \\
\end{array}
\right] 
\nonumber \\
&=& 
U_{23} (\theta_{23}) U_{13} (\theta_{13}) 
U_{12} \left( \frac{\pi}{2} - \theta_{12}, \delta \pm \pi \right) 
\left[
\begin{array}{c}
e^{ i \delta } \nu_{2}  \\
- e^{ - i \delta } \nu_{1} \\
\nu_{3} \\
\end{array}
\right].
\label{flavor-mass-vacuum}
\end{eqnarray}
In eq.~\eqref{flavor-mass-vacuum}, we have used the $U$ matrix of the slightly different convention from that of Particle Data Group (PDG)~\cite{ParticleDataGroup:2022pth}, which we call the SOL convention~\cite{Martinez-Soler:2018lcy}: 
\begin{eqnarray} 
U &=& 
\left[
\begin{array}{ccc}
1 & 0 &  0  \\
0 & e^{ - i \delta} & 0 \\
0 & 0 & e^{ - i \delta} \\
\end{array}
\right] 
U_{\text{\tiny PDG}} 
\left[
\begin{array}{ccc}
1 & 0 &  0  \\
0 & e^{ i \delta} & 0 \\
0 & 0 & e^{ i \delta} \\
\end{array}
\right] 
= 
\left[
\begin{array}{ccc}
1 & 0 &  0  \\
0 & c_{23} & s_{23} \\
0 & - s_{23} & c_{23} \\
\end{array}
\right] 
\left[
\begin{array}{ccc}
c_{13}  & 0 & s_{13} \\
0 & 1 & 0 \\
- s_{13} & 0 & c_{13} \\
\end{array}
\right] 
\left[
\begin{array}{ccc}
c_{12} & s_{12} e^{ i \delta}  &  0  \\
- s_{12} e^{- i \delta} & c_{12} & 0 \\
0 & 0 & 1 \\
\end{array}
\right] 
\nonumber \\
&\equiv& 
U_{23} (\theta_{23}) U_{13} (\theta_{13}) U_{12} (\theta_{12}, \delta),  
\label{U-SOL-def} 
\end{eqnarray}
where $c_{12} \equiv \cos \theta_{12}$, $s_{12} \equiv \sin \theta_{12}$ etc.. 
The reason for our terminology of ``SOL'' is because the CP phase factor $e^{ \pm i \delta}$ is attached to sine of the ``solar angle'' $\theta_{12}$ in the $U$ matrix. Whereas in $U_{\text{\tiny PDG}}$, $e^{ \pm i \delta}$ is attached to $s_{13}$~\cite{ParticleDataGroup:2022pth}. Since the factors in eq.~\eqref{U-SOL-def} sandwiching $U_{\text{\tiny PDG}}$ are the phase factors that can be absorbed into the neutrino states in the left- and right-hand sides of the equation, 
the expressions of the oscillation probabilities are exactly the same as those computed with $U_{\text{\tiny PDG}}$~\cite{Martinez-Soler:2018lcy}. We use the SOL convention $U$ matrix in eq.~\eqref{U-SOL-def} throughout this paper. 

It is important to recognize that eq.~\eqref{flavor-mass-vacuum} implies the symmetry~\cite{Parke:2018shx}. Since rephasing of the states do not affect the observables, the second and third equalities in eq.~\eqref{flavor-mass-vacuum} imply existence of the 1-2 exchange symmetries, invariance under the transformations  
\begin{eqnarray} 
&& 
\text{Symmetry IA-vacuum:}
\hspace{6mm}
m^2_{1} \leftrightarrow m^2_{2}, 
\hspace{8mm}
c_{12} \rightarrow - s_{12},
\hspace{8mm}
s_{12} \rightarrow c_{12}, 
\nonumber \\
&&
\text{Symmetry IB-vacuum:}
\hspace{6mm}
m^2_{1} \leftrightarrow m^2_{2}, 
\hspace{8mm}
c_{12} \leftrightarrow s_{12}, 
\hspace{8mm}
\delta \rightarrow \delta \pm \pi, 
\label{Symmetry-IA-IB-vacuum}
\end{eqnarray} 
where in the first line the possibility of an alternative choice, $c_{12} \rightarrow s_{12}$ and $s_{12} \rightarrow - c_{12}$ ($\theta_{12} \rightarrow \theta_{12} - \frac{\pi}{2}$), is understood. 
In eq.~\eqref{Symmetry-IA-IB-vacuum}, the terms ``Symmetry IA (or IB)-vacuum'' follows our unified classification scheme of the symmetries~\cite{Minakata:2022yvs,Minakata:2021dqh,Minakata:2021goi,Minakata:2022zua}. 

It should not be difficult to recognize that Symmetry IB-vacuum is the ``light-dark'' symmetry mentioned in eq.~\eqref{light-dark}. The readers might wonder why the transformation of $\delta$ is involved in Symmetry IB, but this is the correct symmetry of neutrino oscillation probability in vacuum including the appearance channels. The transformation of $\delta$ is not discussed in the context of the ``light-dark'' symmetry because the survival probability $P(\nu_{e} \rightarrow \nu_{e})$ is free from $\delta$ in vacuum and in matter~\cite{Kuo:1987km,Minakata:1999ze}. 

\section{Reparametrization (Rep) symmetry in matter}
\label{sec:colorful-world}

Now, we enter into the symmetry in neutrino oscillation in matter. We must tell the readers that in matter environment a ``colorful world'' exists for neutrino symmetry. That is, the Rep symmetry in matter is much more profound than that in vacuum. But, why is it so? 

We must first note that introduction of the Wolfenstein matter potential~\cite{Wolfenstein:1977ue}, 
\begin{eqnarray} 
a(x) &=&  
2 \sqrt{2} G_F N_e E \approx 1.52 \times 10^{-4} \left( \frac{Y_e \rho}{\rm g\,cm^{-3}} \right) \left( \frac{E}{\rm GeV} \right) {\rm eV}^2, 
\label{matt-potential}
\end{eqnarray}
into the system completely change the neutrino evolution. In eq.~\eqref{matt-potential}, $G_F$ is the Fermi constant, $N_e$ and $N_n$ are the electron and neutron number densities in matter. $\rho$ and $Y_e$ denote, respectively, the matter density and number of electrons per nucleon in matter. The matter potential $a$ has the dimension of energy squared, the same dimension as $\Delta m^2$'s, but they come-in into the system with the different flavor dependences. It makes the neutrino evolution much more dynamical than that in vacuum. 

The most spectacular case of the matter effect on neutrinos occurs in the solar neutrinos which are produced in the central region in the Sun with energies larger than a few MeV. Due to high matter density $\rho \sim 100$\,gcm$^{-3}$, $\nu_{e}$ is produced dominantly as the mass eigenstate $\nu_{2}$ in matter in the two-flavor approximation, and it evolves adiabatically by passing through the level crossing region. The behavior of the mixing angle in matter signals the resonant behavior. Then, neutrino leaves the Sun as the vacuum $\nu_{2}$ state, in which the $\nu_{e}$ component is only a small fraction $\sin^2 \theta_{12} \simeq 0.3$, the adiabatic flavor conversion~\cite{Mikheyev:1985zog}. It provides so called the ``large mixing angle'' (LMA) MSW solution of the solar neutrino problem~\cite{Wolfenstein:1977ue,Mikheyev:1985zog,Maltoni:2015kca}. 

On the other hand, in the kinematical regions relevant for (most of) the terrestrial accelerator and atmospheric neutrino experiments, enhancement of the neutrino oscillations occurs~\cite{Barger:1980tf} due to the Earth matter effect, but without passage of the neutrino states through the level crossing region. One can see even in perturbative manner that an atmospheric-resonance-like peak is developed when the higher order effect of ``large $\theta_{13}$'' is taken into account~\cite{Minakata:2009sr,Asano:2011nj}. It is the oscillations enhanced by the matter effect, and hence it may or may not be called as the {\it bona fide} resonance. The complexity of inter-relationship of various features of the neutrino oscillations and adiabatic conversion in matter may be understood most clearly by using the graphical representations based on an analogy with the spin precession, see e.g., ref.~\cite{Smirnov:2016xzf}. The matter enhanced oscillations corresponds to Fig.~11c in ref.~\cite{Smirnov:2016xzf}, which is labelled the ``resonance enhancement of oscillations''. Following this terminology, we use ``resonance'' as the collective nomenclature for the matter enhanced oscillations even though the maximal matter-affected mixing angle may not be reached. 

\begin{figure}[h!]
\begin{center}
\vspace{4mm}
\includegraphics[width=0.6\textwidth]{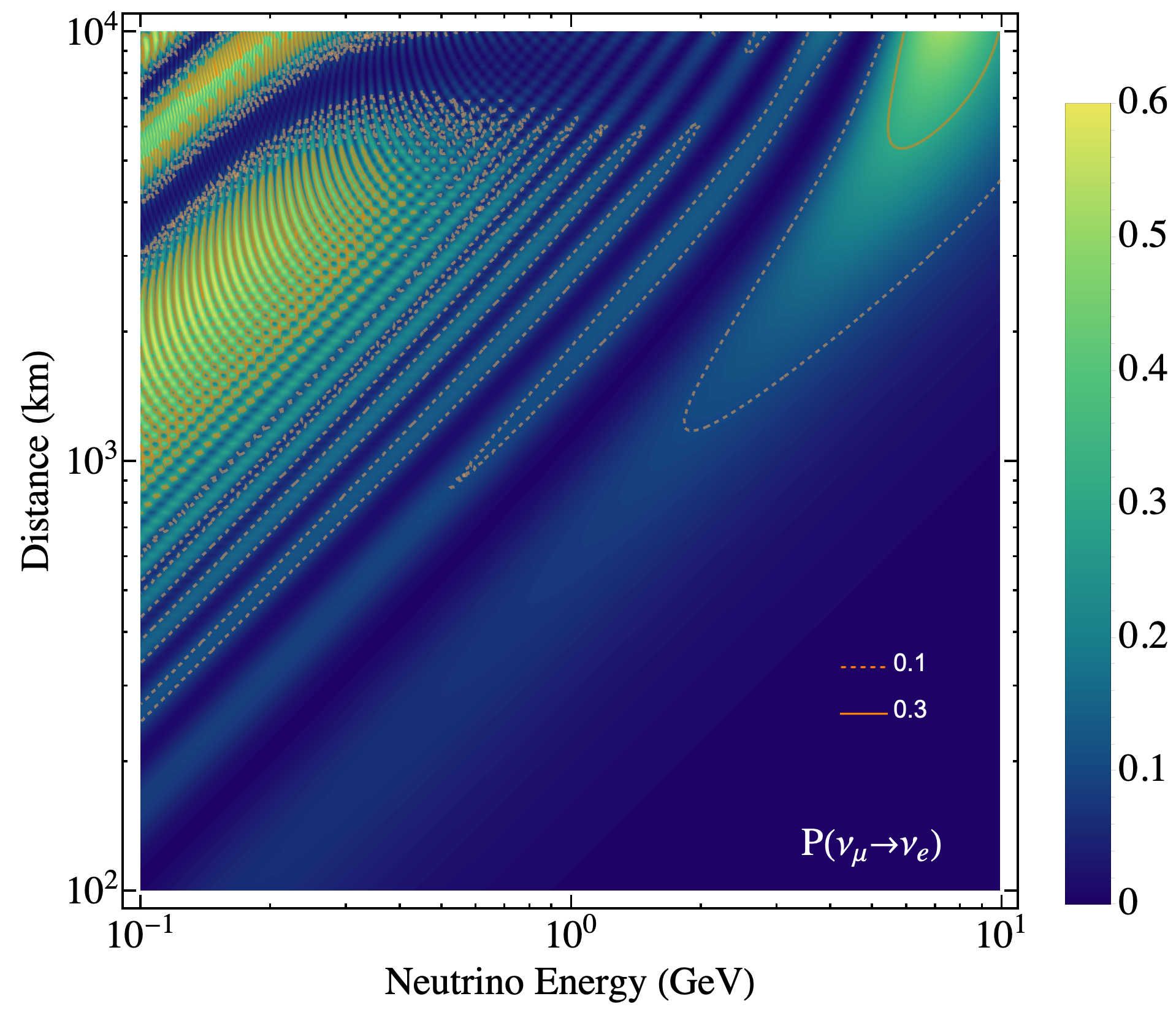}
\end{center}
\vspace{-2mm}
\caption{The equi-probability contour of $P(\nu_{\mu} \rightarrow \nu_{e})$ is presented~\cite{Minakata:2019gyw} in region of energy-baseline that covers the atmospheric neutrino observation by Super-Kamiokande. The two enhanced regions correspond, respectively, to the solar-scale and atmospheric-scale resonances, see the main text. The matter density is taken to be a constant, $\rho = 4.0~\text{g/cm}^3$, which gives only a bold approximation to the Earth matter density. 
} 
\vspace{-2mm}
\label{fig:global-oscillation}
\end{figure}

Our theoretical discussion assumes tentatively the world that can be explored by the terrestrial neutrino experiments. Then, the resonance enhancement of oscillations plays a role to enrich the features of neutrino evolution in matter. 
To appeal the readers' intuition of these enhancements, we depict in Fig.~\ref{fig:global-oscillation} the equi-probability contour of $P(\nu_{\mu} \rightarrow \nu_{e})$~\cite{Minakata:2019gyw} in region of energy-baseline that roughly covers the Super-Kamiokande's atmospheric neutrino observation, 0.1 GeV$\lsim E \lsim 10$ GeV, see Fig.~3 in ref.~\cite{Super-Kamiokande:2017yvm}. The enhanced regions at around $E \sim 200$ MeV, $L \sim 2000$ km and at $E \sim 8$ GeV, $L \sim 10^{4}$ km  correspond, respectively, to the solar-scale and atmospheric-scale enhanced oscillations. The features depicted in Fig.~\ref{fig:global-oscillation} is sufficiently rich to stimulate the various formulations of oscillations to accommodate them, which entail the profound symmetries. They include symmetries not only of the 1-2 state exchange type, but also the 1-3 exchange one~\cite{Minakata:2022yvs,Minakata:2021dqh,Minakata:2021goi}.

\section{Symmetry Finder (SF) and a status summary of the SF symmetries} 
\label{sec:SF} 

To make a systematic search for the Rep symmetry in neutrino oscillation in matter, we have introduced a powerful machinery called ``Symmetry Finder'' (SF)~\cite{Minakata:2022yvs,Minakata:2021dqh,Minakata:2021goi,Minakata:2022zua}, following the spirit of the scheme in vacuum expressed in eq.~\eqref{flavor-mass-vacuum}. Please be patient: We will give a compact summary of the SF formalism by taking SRP and DMP as the concrete frameworks for our comparative study of their 1-2 symmetries. 

Here is a brief summary of symmetries in neutrino oscillation in matter to date. In this paper we restrict our discussions on the Rep symmetry into the ones in the $\nu$SM context. For the symmetries in the extended frameworks with nonunitarity, which is a low energy manifestation of new physics beyond the $\nu$SM~\cite{Antusch:2006vwa}, see Refs.~\cite{Minakata:2022yvs,Minakata:2022zua}. 
Thus far, the SF symmetry search was undertaken in the SRP~\cite{Martinez-Soler:2019nhb}, the helio-perturbation~\cite{Minakata:2015gra}, and the DMP perturbation theories~\cite{Denton:2016wmg}, which entailed the following results: 
\begin{itemize}

\item 
Eight symmetries of the 1-2 state exchange type in the SRP~\cite{Minakata:2022yvs} and DMP~\cite{Minakata:2021dqh}.

\item 
Sixteen symmetries of the 1-3 state exchange type in the helio-perturbation theory~\cite{Minakata:2021goi}.\footnote{
The terminology ``1-3 exchange symmetry'' is a symbolic one. In the normal (inverted) mass ordering $\lambda_{3} > \lambda_{2}$ ($\lambda_{1} > \lambda_{3}$) are the two eigenvalues which have the level crossing~\cite{Minakata:2015gra}. In refs.~\cite{Minakata:2021goi,Minakata:2015gra}, the unified notation for the level crossing states $\nu_{-}$ and $\nu_{+}$ is introduced, which is valid for the both mass orderings.  }

\end{itemize}
In agreement with the aforementioned expectation for enriched symmetry in matter, we have obtained quite a large number of the Rep symmetries. Among them, the SRP and DMP symmetries will be discussed and summarized in Table~\ref{tab:SRP-symmetry} and Table~\ref{tab:DMP-symmetry}, respectively, in section~\ref{sec:comparative}. 

We just mention the important characteristics of the SF symmetries, see refs.~\cite{Minakata:2022yvs,Minakata:2021dqh,Minakata:2021goi,Minakata:2022zua} for details. That is, one can prove that the SF symmetry is a Hamiltonian symmetry, which means that the flavor basis Hamiltonian is kept invariant up to the rephasing factor (which does not affect the observables) under the symmetry transformations. It implies: 
\begin{itemize}

\item 
The symmetry holds to all orders in perturbation theory, even though our SF treatment itself is valid only to the first order. At the same time, the Hamiltonian proof ensures that the symmetry holds even in a varying matter-density environment. 

\end{itemize}

\subsection{Globally valid vs. locally valid frameworks} 
\label{sec:globally-valid} 

It is instructive to introduce a classification of the perturbative frameworks of neutrino oscillation in matter~\cite{Minakata:2022yvs}. 
By ``globally-valid'' framework we mean, roughly speaking, that it is valid throughout the region depicted in Fig.~\ref{fig:global-oscillation}. It contains the kinematical regions covered by the most, if not all, of the terrestrial experiments ongoing or planned. There exist only the two known examples of the globally-valid frameworks, DMP~\cite{Denton:2016wmg} and the Agarwalla {\it et al.} (AKT)~\cite{Agarwalla:2013tza} perturbation theories. In fact, the region of validity of these theories is likely to extend to much higher energies which is explored e.g., by IceCube-DeepCore~\cite{IceCube:2020tka} and/or KM3NeT/ORCA~\cite{KM3NeT:2021ozk}. For the related discussions on the region of validity, see e.g., refs.~\cite{Minakata:2021nii,Parke:2019jyu}. 
On the other hand, the examples of the locally-valid theory include: SRP valid at around the solar resonance~\cite{Martinez-Soler:2019nhb}, and the helio-perturbation theory valid in the atmospheric resonance region~\cite{Minakata:2015gra}. See e.g., refs.~\cite{Arafune:1996bt,Cervera:2000kp,Freund:2001pn,Akhmedov:2004ny} for the earlier versions of the atmospheric-resonance perturbation theory. 

For reasons we will explain in the next section~\ref{sec:new-picture}, we will undertake a comparative study of the two 1-2 exchange symmetries, one in locally-valid SRP and the other in globally-valid DMP. It may illuminate how the unique physical object, solar resonance, is described by the locally- and globally-valid frameworks, making the exercise an interesting task by itself.

\section{A new picture {\em ``one-resonance - one-symmetry''} and the framework dependence of the symmetry}
\label{sec:new-picture} 

We feel it instructive to present a bold ansatz, or an imaginative view, to understand the current figure of the Rep symmetry in neutrino oscillation in matter~\cite{Minakata:2022yvs}: 
\begin{itemize} 
\item 
Occurrence of the symmetry takes place in correlation with the regions of enhanced oscillation, at around the solar and atmospheric resonances. 

\end{itemize}
The picture is in fact born out from the 1-2 and 1-3 exchange symmetries in SRP~\cite{Minakata:2022yvs} and the helio-perturbation theory~\cite{Minakata:2021goi}, respectively, obtained in the above mentioned two locally-valid frameworks. The states exchanged in these symmetries are the ones that participate the level crossing in the solar and atmospheric resonances, the major players in the respective regions. For {\em one}~region of enhancement of oscillation, there exists {\em one}~symmetry of the particular state exchange type that corresponds to the level crossing. This picture has a predictive power on where the symmetry resides and which state exchange is involved in the symmetry. The feature makes this picture attractive. 

\subsection{Framework dependence of the reparametrization (Rep) symmetry}
\label{sec:framework-dep} 

The existence of the Rep symmetry implies that there is another way of parametrizing the equivalent solution of the theory. In our current formulation of neutrino evolution, the original theory and its symmetry copy are defined in a chosen particular framework of neutrino oscillation, and hence the Rep symmetry is framework dependent by construction. In our SF machinery, therefore, we may not reach a ``general symmetry'' which can be extracted from a given Hamiltonian itself, the problem mentioned in ref.~\cite{Minakata:2022yvs}. See ref.~\cite{Denton:2021vtf} for a possibly relevant approach to this problem. 

However, our ``one-resonance - one-symmetry'' picture may imply a departure from this complete framework dependence. If each symmetry is associated with the corresponding resonance enhancement, a physical phenomenon that exists in any frameworks for describing the enhancement, the issue of framework dependence is no longer real. In that case, the Rep symmetry reflects the physical feature of the neutrino evolution, not an artifact of the particular framework taken. 

We want to confirm, or refute, the one-resonance - one-symmetry picture against the existing Rep symmetry ``data''. As we stated in section~\ref{sec:SF}, SRP and DMP both have their own 1-2 exchange symmetry. Then, we can ask: Are these two 1-2 symmetries really identical to each other? If the symmetry is associated with the resonance, not their frameworks, the symmetry possessed by these two theories must be identical. This consideration naturally leads us to a comparative study of the SRP and DMP 1-2 exchange symmetries to which we are now ready to enter. 

\section{Comparative study of the SRP and DMP symmetries} 
\label{sec:comparative} 

We investigate the problem of identity of the two 1-2 state exchange symmetries, one in SRP and the other in DMP. We aim at superseding the original treatments given in refs.~\cite{Minakata:2022yvs,Minakata:2021dqh}. The reasons for our revised treatment are that the previous ones lacked a sharp focus on the meaning of this identity, and the logic behind seeking the identical symmetry in the two different theories (with the substantially different SF equations) was not transparent. Interestingly, our investigation here will reveal the (almost) identity between the SRP and DMP 1-2 symmetries, but in a highly nontrivial way. 

A few words on the SRP theory~\cite{Martinez-Soler:2019nhb}. It is the locally-valid theory at around the solar-scale enhancement, neutrino energy $E=( 1 - 5 ) \times 100$ MeV and baseline $L= ( 1 - 10 ) \times 1000$ km. See Fig.~\ref{fig:global-oscillation}. In this region, the matter potential $a$ defined in eq.~\eqref{matt-potential} is comparable in size to the vacuum effect represented by $\Delta m^2_{21}$, $a/\Delta m^2_{21} \simeq 0.61$ for $\rho=3.0~\text{g/cm}^3$ and $E=200$ MeV. We use the uniform matter-density approximation in this paper. 

\subsection{Flavor state vs. energy eigenstate in SRP and DMP}
\label{sec:flavor-mass-SRP-DMP} 

Since the SRP (solar-resonance perturbation) theory is reviewed to a reasonable depth and its SF formulation is set up in ref.~\cite{Minakata:2022yvs}, we just start by recollecting the relevant formulas. We utilize the $V$ matrix formalism~\cite{Minakata:1998bf} to express the flavor state by the mass eigenstate as $\nu_{\alpha} = V_{\alpha, i} \check{\nu}_{i}$, where $\alpha$ runs over $e, \mu, \tau$, and $i=1,2,3$. 
Using the result of explicit computation done in ref.~\cite{Minakata:2022yvs}, the flavor state can be written to first order in the SRP theory by using the $V$ matrix and the mass eigenstate, in the SOL convention, as 
\begin{eqnarray} 
&& 
\left[
\begin{array}{c}
\nu_{e} \\
\nu_{\mu} \\
\nu_{\tau} \\
\end{array}
\right] 
= 
U_{23} (\theta_{23}) U_{13} (\theta_{13}) U_{12} (\varphi, \delta) 
\biggl\{
1 + \mathcal{W}_{ \text{SRP} }^{(1)} ( \theta_{13}, \varphi, \delta; \lambda_{1}, \lambda_{2} )
\biggr\} 
\left[
\begin{array}{c}
\nu_{1} \\
\nu_{2} \\
\nu_{3} \\
\end{array}
\right], 
\label{flavor-state-SRP}
\end{eqnarray}
where $\mathcal{W}_{ \text{SRP} }^{(1)} ( \theta_{13}, \varphi, \delta; \lambda_{1}, \lambda_{2} )$ is defined by 
\begin{eqnarray} 
&&
\mathcal{W}_{ \text{SRP} }^{(1)} ( \theta_{13}, \varphi, \delta; \lambda_{1}, \lambda_{2} ) 
\equiv 
c_{13} s_{13} 
\left[
\begin{array}{ccc}
0 & 0 & c_{\varphi} \frac{ a }{ \lambda_3 - \lambda_1 } \\
0 & 0 & s_{\varphi} e^{ - i \delta} \frac{ a }{ \lambda_3 - \lambda_2 } \\
- c_{\varphi} \frac{ a }{ \lambda_3 - \lambda_1 } & 
- s_{\varphi} e^{ i \delta} \frac{ a }{ \lambda_3 - \lambda_2 } & 0 \\
\end{array}
\right]. 
\label{W-SM-SRP1st}
\end{eqnarray} 

In DMP, the similar expression of the flavor state with the use of the $V$ matrix is given to first order in the DMP expansion as~\cite{Minakata:2021dqh}
\begin{eqnarray} 
\left[
\begin{array}{c}
\nu_{e} \\
\nu_{\mu} \\
\nu_{\tau} \\
\end{array}
\right] 
&=& 
U_{23} (\theta_{23}) U_{13} (\phi) U_{12} (\psi, \delta)
\left\{ 
1 + 
\mathcal{W}_{ \text{DMP} }^{(1)} 
( \theta_{12}, \theta_{13}, \delta, \phi, \psi; \lambda_{1}, \lambda_{2} ) 
\right\} 
\left[
\begin{array}{c}
\nu_{1} \\
\nu_{2} \\
\nu_{3} \\
\end{array}
\right],
\label{flavor-state-DMP}
\end{eqnarray}
where $\mathcal{W}_{ \text{DMP} }^{(1)} ( \theta_{12}, \theta_{13}, \delta, \phi, \psi; \lambda_{1}, \lambda_{2} )$ is given by 
\begin{eqnarray} 
&&
\hspace{-3mm}
\mathcal{W}_{ \text{DMP} }^{(1)} ( \theta_{12}, \theta_{13}, \delta, \phi, \psi; \lambda_{1}, \lambda_{2} ) 
= 
\epsilon c_{12} s_{12} \sin ( \phi - \theta_{13} ) 
\left[
\begin{array}{ccc}
0 & 0 & - s_\psi 
\frac{ \Delta m^2_{ \text{ren} } }{ \lambda_{3} - \lambda_{1} } \\
0 & 0 & c_\psi e^{ - i \delta} 
\frac{ \Delta m^2_{ \text{ren} } }{ \lambda_{3} - \lambda_{2} } \\
s_\psi 
\frac{ \Delta m^2_{ \text{ren} } }{ \lambda_{3} - \lambda_{1} } & 
- c_\psi e^{ i \delta} 
\frac{ \Delta m^2_{ \text{ren} } }{ \lambda_{3} - \lambda_{2} } & 0 \\
\end{array}
\right]. ~~~~
\nonumber \\
\label{W-SM-DMP1st}
\end{eqnarray}

In eqs.~\eqref{W-SM-SRP1st} and \eqref{W-SM-DMP1st}
$\lambda_{i}$ are the eigenvalues of the Hamiltonian times $2E$ in the leading order in SRP and DMP perturbation theories, respectively. The explicit expressions of $\lambda_{i}$ in SRP will be given in eq.~\eqref{eigenvalues-SRP}, but the ones in DMP are less simple as shown in ref.~\cite{Denton:2016wmg}. 
In SRP $\varphi$ is the matter-dressed $\theta_{12}$, and in DMP $\psi$ and $\phi$ denote, respectively, $\theta_{12}$ and $\theta_{13}$ in matter. $c_{\psi}$ and $s_{\psi}$ are shorthand notations for $\cos \psi$ and $\sin \psi$, respectively. 
As in the helio-perturbation theory~\cite{Minakata:2015gra}, $\epsilon \equiv \Delta m^2_{21} / \Delta m^2_{ \text{ren} }$  is the unique expansion parameter in the DMP perturbation theory~\cite{Denton:2016wmg}, where its denominator is defined as $\Delta m^2_{ \text{ren} } \equiv \Delta m^2_{31} - s^2_{12} \Delta m^2_{21}$. We need a careful discussion on the effective expansion parameter in SRP, $A_{ \text{exp} } = c_{13} s_{13} (a / \Delta m^2_{31}) \simeq 2.8 \times 10^{-3}$~\cite{Martinez-Soler:2019nhb} at the above values of $\rho$ and $E$. It is a ``framework generated'' expansion parameter, meaning that its smallness is partly due to the propagator suppression as one can observe in eq.~\eqref{W-SM-SRP1st}. See refs.~\cite{Martinez-Soler:2019nhb,Minakata:2022yvs} for more details. 

\subsection{Symmetry Finder (SF) equation in SRP and DMP}
\label{sec:SF-SRP-DMP} 

Now let us write down the SF equation in the both SRP and DMP perturbation theories. To prepare the flavor states~\eqref{flavor-state-SRP} and \eqref{flavor-state-DMP} in physically equivalent but slightly extended forms, we introduce the flavor-state rephasing matrix $F$ and the generalized 1-2 state exchange matrix $R$, which are defined by 
\begin{eqnarray} 
&&
F \equiv 
\left[
\begin{array}{ccc}
e^{ i \tau } & 0 & 0 \\
0 & e^{ i \sigma } & 0 \\
0 & 0 & 1 \\
\end{array}
\right], 
\hspace{8mm}
R \equiv 
\left[
\begin{array}{ccc}
0 & - e^{ i ( \delta + \alpha) } & 0 \\
e^{ - i ( \delta + \beta) } & 0 & 0 \\
0 & 0 & 1 \\
\end{array}
\right]. 
\label{F-R-SRP-def}
\end{eqnarray}
The matrices $F$ and $R$ in eq.~\eqref{F-R-SRP-def} take the nonvanishing and non-unity elements only in the 1-2 sub-sector because we restrict ourselves into the 1-2 state exchange symmetry in the both theories. 
Then, the SF equation reads in SRP: 
\begin{eqnarray} 
&& 
F \left[
\begin{array}{c}
\nu_{e} \\
\nu_{\mu} \\
\nu_{\tau} \\
\end{array}
\right] 
= 
F U_{23} (\theta_{23}) F^{\dagger}
F U_{13} (\theta_{13}) F^{\dagger} 
F U_{12} (\varphi, \delta) 
R^{\dagger} R 
\biggl\{
1 + \mathcal{W}_{ \text{SRP} }^{(1)} ( \theta_{13}, \varphi, \delta; \lambda_{1}, \lambda_{2} ) 
\biggr\} 
R^{\dagger} R 
\left[
\begin{array}{c}
\nu_{1} \\
\nu_{2} \\
\nu_{3} \\
\end{array}
\right] 
\nonumber \\
&=& 
U_{23} (\theta_{23}^{\prime}) U_{13} (\theta_{13}^{\prime}) 
U_{12} ( \varphi^{\prime}, \delta + \xi) 
%
\biggl\{
1 + \mathcal{W}_{ \text{SRP} }^{(1)} ( \theta_{13}^{\prime}, \varphi^{\prime}, \delta + \xi; \lambda_{2}, \lambda_{1} ) 
\biggr\} 
\left[
\begin{array}{c}
- e^{ i ( \delta + \alpha) } \nu_{2} \\
e^{ - i ( \delta + \beta) } \nu_{1} \\
\nu_{3} \\
\end{array}
\right], 
\label{SFeq-SRP} 
\end{eqnarray}
and in DMP: 
\begin{eqnarray} 
&& 
F \left[
\begin{array}{c}
\nu_{e} \\
\nu_{\mu} \\
\nu_{\tau} \\
\end{array}
\right] 
= 
F U_{23} (\theta_{23}) F^{\dagger}
F U_{13} (\phi) F^{\dagger} 
F U_{12} (\psi, \delta) 
R^{\dagger} R 
\biggl\{
1 + 
\mathcal{W}_{ \text{DMP} }^{(1)} 
( \theta_{12}, \theta_{13}, \delta, \phi, \psi; \lambda_{1}, \lambda_{2} ) 
\biggr\} 
R^{\dagger} R 
\left[
\begin{array}{c}
\nu_{1} \\
\nu_{2} \\
\nu_{3} \\
\end{array}
\right] 
\nonumber \\
&=& 
U_{23} (\theta_{23}^{\prime}) U_{13} (\phi^{\prime}) 
U_{12} ( \psi^{\prime}, \delta + \xi) 
%
\biggl\{
1 + \mathcal{W}_{ \text{DMP} }^{(1)} 
( \theta_{12}^{\prime}, \theta_{13}^{\prime}, \delta + \xi, \phi^{\prime}, \psi^{\prime}; \lambda_{2}, \lambda_{1} ) 
\biggr\} 
\left[
\begin{array}{c}
- e^{ i ( \delta + \alpha) } \nu_{2} \\
e^{ - i ( \delta + \beta) } \nu_{1} \\
\nu_{3} \\
\end{array}
\right],
\label{SFeq-DMP} 
\end{eqnarray}
where we have denoted the transformed $\delta$ as $\delta^{\prime} = \delta + \xi$. 

Notice that the first line in the right-hand side of eq.~\eqref{SFeq-SRP} is physically equivalent to the state~\eqref{flavor-state-SRP} because $R^{\dagger} R=1$ and the rephasing $F$ matrix does not alter the physical content of the state. Whereas, the second line of eq.~\eqref{SFeq-SRP} is the expression of the same flavor state by the transformed mass eigenstate with the 1-2 state exchange and the transformed $\nu$SM variables. The discussion of the SF equation in DMP is completely parallel. If the SF equation has a solution, it implies existence of a symmetry, in perfect parallelism with the discussion in vacuum in section~\ref{sec:systematic-way}. 

In fact, the introduction of the $F$ matrix turns out to be important, by which the vacuum angles $\theta_{23}$ and $\theta_{13}$ in the SRP ($\phi$ in DMP) transform, in general, resulting in the enriched symmetry list. 

\subsection{Solutions of the SF equation: SRP vs. DMP}
\label{sec:SF-solution-SRP-DMP} 

The SF equation can be decomposed into the three parts, (1) the overall factor, (2) the zeroth-order and (3) the first-order pieces. They will be presented in order below, and (2) and (3) are denoted as the first and the second conditions, respectively. Let us focus on the SRP theory first. The overall condition reads 
\begin{eqnarray} 
F U_{23} (\theta_{23}) F^{\dagger} F U_{13} (\theta_{13}) F^{\dagger} 
&=&
U_{23} (\theta_{23}^{\prime}) U_{13} (\theta_{13}^{\prime}), 
\label{overall-condition}
\end{eqnarray}
which can be solved as $\sigma = \pm \pi$ and $\tau = \pm \pi$ under the ansatz $s_{23} e^{ i \sigma } = s_{23}^{\prime}$ and $s_{13} e^{ i \tau } = s_{13}^{\prime}$, because apparently we have no other choice, assuming that $s_{ij}^{\prime}$ are real numbers, within the present SF formalism.

The first and the second conditions read: 
\begin{eqnarray} 
%
F U_{12} (\varphi, \delta) R^{\dagger} 
&=&  
U_{12} (\varphi^{\prime}, \delta + \xi), 
\nonumber \\
R 
\mathcal{W}_{ \text{SRP} }^{(1)} ( \theta_{13}, \varphi, \delta; \lambda_{1}, \lambda_{2} ) 
R^{\dagger} 
&=& 
\mathcal{W}_{ \text{SRP} }^{(1)} ( \theta_{13}^{\prime}, \varphi^{\prime}, \delta+ \xi; \lambda_{2}, \lambda_{1} ). 
\label{1st-2nd-conditions}
\end{eqnarray}
The first condition can be reduced to the forms, 
$c_{\varphi^{\prime}} = - s_{\varphi} e^{ - i ( \alpha - \tau ) } = - s_{\varphi} e^{ i ( \beta + \sigma ) }$, and $s_{\varphi^{\prime}} = c_{\varphi} e^{ i ( \beta + \tau - \xi ) } = c_{\varphi} e^{ - i ( \alpha - \sigma - \xi ) }$. We note that these equations together with the above restrictions of $\tau$ and $\sigma$ being integer multiples of $\pi$, the phases $\xi$, $\alpha$, and $\beta$ must also be integer multiples of $\pi$~\cite{Minakata:2022yvs,Minakata:2021dqh}. 
By changing $\varphi$ to $\psi$, we obtain the first condition in DMP. The solutions of the first condition, which is common to SRP and DMP, are tabulated in Table~\ref{tab:SF-solutions}, establishing the classification schemes of the Rep symmetries. Interestingly, Table~\ref{tab:SF-solutions} is universally valid~\cite{Minakata:2022yvs} not only in SRP and DMP, both with the 1-2 symmetry, but also in the helio-perturbation theory whose symmetry is the 1-3 exchange type~\cite{Minakata:2021goi}. 

\begin{table}[h!]
\vglue -0.2cm
\begin{center}
\caption{ 
The solutions of the first condition in SRP and DMP. The labels ``upper'' and ``lower'' imply the upper and lower sign in the corresponding columns in Table~\ref{tab:SRP-symmetry} and Table~\ref{tab:DMP-symmetry}. 
}
\label{tab:SF-solutions}
\vglue 0.2cm
\begin{tabular}{c|c|c}
\hline 
Symmetry & 
$\tau, \sigma, \xi$ & 
$\alpha, \beta$
\\
\hline 
\hline 
Symmetry IA & 
$\tau = \sigma = 0$, $\xi = 0$ & 
$\alpha = \beta = 0$ (upper) \\ 
& &
$\alpha = \pi, \beta = - \pi$ (lower)  \\
\hline
Symmetry IB & 
$\tau = \sigma = 0$, $\xi = \pi$ & 
$\alpha = \pi, \beta = - \pi$ (upper) \\
& & $\alpha = \beta = 0$ (lower) \\
\hline 
Symmetry IIA & 
$\tau = 0, \sigma = - \pi$, $\xi = 0$ & 
$\alpha = \pi, \beta = 0$ (upper) \\
& & $\alpha = 0, \beta = \pi$ (lower)  \\
\hline 
Symmetry IIB & 
$\tau = 0, \sigma = - \pi$, $\xi = \pi$ & 
$\alpha = 0, \beta = \pi$ (upper) \\ 
& & $\alpha = \pi, \beta = 0$ (lower) \\
\hline 
Symmetry IIIA & 
$\tau = \pi, \sigma = 0$, $\xi = 0$ & 
$\alpha = 0, \beta = \pi$ (upper) \\ 
& & $\alpha = \pi, \beta = 0$ (lower) \\
\hline 
Symmetry IIIB & 
$\tau = \pi, \sigma = 0$, $\xi = \pi$ & 
$\alpha = \pi, \beta = 0$ (upper)  \\ 
 & & 
$\alpha = 0, \beta = \pi$ (lower)   \\
\hline 
Symmetry IVA & 
$\tau = \sigma = \pi$, $\xi = 0$ & 
$\alpha = \pi, \beta = - \pi$ (upper) \\ 
& &
$\alpha = \beta = 0$ (lower)  \\
\hline 
Symmetry IVB & 
$\tau = \sigma = \pi$, $\xi = \pi$ & 
$\alpha = \beta = 0$ (upper) \\ 
& &
$\alpha = \pi, \beta = - \pi$ (lower)  \\
\hline 
\end{tabular}
\end{center}
\vglue -0.4cm 
\end{table}

\subsection{Second conditions in SRP and DMP are {\em not} so similar} 
\label{sec:2nd-condition-SRP-DMP} 

Despite the above parallelism, SRP and DMP are completely different perturbation theory by having the qualitatively different expansion parameters, as briefly explained in section~\ref{sec:flavor-mass-SRP-DMP}. Reflecting this difference the second conditions differ between these two theories. It takes the following form in SRP :  
\begin{eqnarray} 
&&
c_{13} s_{13}
\left[
\begin{array}{ccc}
0 & 0 & 
- e^{ i \alpha } s_{\varphi} \frac{ a }{ \lambda_3 - \lambda_2 } \\
0 & 0 & 
e^{ - i ( \delta + \beta) } c_{\varphi} \frac{ a }{ \lambda_3 - \lambda_1 } \\
e^{ - i \alpha } s_{\varphi} \frac{ a }{ \lambda_3 - \lambda_2 } & 
- e^{ i ( \delta + \beta) } c_{\varphi} \frac{ a }{ \lambda_3 - \lambda_1 } & 0 \\
\end{array}
\right] 
\nonumber \\
&=& 
c_{13}^{\prime} s_{13}^{\prime}
\left[
\begin{array}{ccc}
0 & 0 & c_{\varphi}^{\prime} \frac{ a }{ \lambda_3 - \lambda_2 } \\
0 & 0 & e^{ - i ( \delta + \xi ) } s_{\varphi}^{\prime} \frac{ a }{ \lambda_3 - \lambda_1 } \\
- c_{\varphi}^{\prime} \frac{ a }{ \lambda_3 - \lambda_2 } & 
- e^{ i ( \delta + \xi ) } s_{\varphi}^{\prime} \frac{ a }{ \lambda_3 - \lambda_1 } & 0 \\
\end{array}
\right], 
\label{2nd-condition-SRP}
\end{eqnarray}
and in DMP:  
\begin{eqnarray} 
&&
\epsilon c_{12} s_{12} \sin ( \phi - \theta_{13} )
\left[
\begin{array}{ccc}
0 & 0 & - c_\psi e^{ i \alpha } 
\frac{ \Delta m^2_{ \text{ren} } }{ \lambda_{3} - \lambda_{2} } \\
0 & 0 & - s_\psi e^{ - i ( \delta + \beta) } 
\frac{ \Delta m^2_{ \text{ren} } }{ \lambda_{3} - \lambda_{1} }  \\
c_\psi e^{ - i \alpha } 
\frac{ \Delta m^2_{ \text{ren} } }{ \lambda_{3} - \lambda_{2} } & 
s_\psi e^{ i ( \delta + \beta) } 
\frac{ \Delta m^2_{ \text{ren} } }{ \lambda_{3} - \lambda_{1} } & 0 \\
\end{array}
\right] 
\nonumber \\
&=&
\epsilon c_{12}^{\prime} s_{12}^{\prime} \sin ( \phi^{\prime} - \theta_{13}^{\prime} )
\left[
\begin{array}{ccc}
0 & 0 & - s_\psi^{\prime}
\frac{ \Delta m^2_{ \text{ren} } }{ \lambda_{3} - \lambda_{2} } \\
0 & 0 & c_\psi^{\prime} e^{ - i ( \delta + \xi) } 
\frac{ \Delta m^2_{ \text{ren} } }{ \lambda_{3} - \lambda_{1} } \\
s_\psi^{\prime}
\frac{ \Delta m^2_{ \text{ren} } }{ \lambda_{3} - \lambda_{2} } & 
- c_\psi^{\prime} e^{ i ( \delta + \xi) } 
\frac{ \Delta m^2_{ \text{ren} } }{ \lambda_{3} - \lambda_{1} } & 0 \\
\end{array}
\right].
\label{2nd-condition-DMP} 
\end{eqnarray}
This difference between the second conditions of eqs.~\eqref{2nd-condition-SRP} and~\eqref{2nd-condition-DMP} makes understanding of the ``identity'' of the SRP and DMP 1-2 symmetries quite nontrivial. 

Despite the identical matrix stricture having zeros in the same entries in common, 
there exist the two important differences between eqs.~\eqref{2nd-condition-SRP} and~\eqref{2nd-condition-DMP}: 
(1) The pre-factor includes $\theta_{12}$ in DMP, but not in SRP. 
(2) To transform the matrix part of the SRP second condition to that of DMP, the following nontrivial transformations are necessary: $s_{\varphi} \rightarrow c_{\psi}, c_{\varphi} \rightarrow - s_{\psi}, a \rightarrow \Delta m^2_{ \text{ren} }$. The $\varphi$ to $\psi$ transformation is the issue, because neither 
$s_{\varphi} \rightarrow c_{\varphi}, c_{\varphi} \rightarrow - s_{\varphi}$ nor its $\psi$ version is not always the symmetry in SRP, nor in DMP, as one can confirm in Table~\ref{tab:SRP-symmetry} and Table~\ref{tab:DMP-symmetry}, respectively. 

Notice that the difference between $\sin ( \phi - \theta_{13} )$ and $c_{13} s_{13}$ in the pre-factors is not essential because the sign flip or non-flip of $\theta_{13}$ is the only concern to us. When $\theta_{13}$ flips sign, sign flip of $\phi$ is enforced in DMP because $\sin 2\phi \propto \sin 2\theta_{13}$~\cite{Denton:2016wmg}. Similarly, a change of $a$ to $\Delta m^2_{ \text{ren} }$ does not affect our symmetry discussion because it is an overall factor in the second condition. Nonetheless, the $a$ to $\Delta m^2_{ \text{ren} }$ change required to make the SRP second condition to DMP's (the matrix parts only) testifies clearly that the SRP and DMP perturbation theories are completely different ones from each other. 
 
\subsection{SRP vs. DMP symmetries: Why are they so similar?}
\label{sec:why-similar} 

We must also observe that when the second conditions~\eqref{2nd-condition-SRP} and~\eqref{2nd-condition-DMP} are solved as a whole for each given solution of the first condition in SRP and DMP, respectively, we obtain the essentially identical 1-2 symmetries Symmetry Z-SRP and Symmetry Z-DMP for all Z=IA, IB, $\cdot \cdot \cdot $, IVB, despite the marked differences between the two theories. See Table~\ref{tab:SRP-symmetry} for SRP and Table~\ref{tab:DMP-symmetry} for DMP. The minor difference exists in the matter undressing of $\theta_{13}$ in SRP, as opposed to $\phi$, the matter-dressed $\theta_{13}$ in DMP. The matter undressed $\theta_{13}$ is sensible in the solar-resonance region, which is away from the atmospheric resonance. In fact, $\phi \approx \theta_{13}$ in a good approximation at the solar resonance, see Fig.~1 in ref.~\cite{Denton:2016wmg}. Notice again that in DMP $\phi$ sign flip is always correlated with $\theta_{13}$ sign flip. 

\begin{table}[h!]
\begin{center}
\caption{All the reparametrization (Rep) symmetries of the 1-2 state exchange type found in the solar-resonance perturbation (SRP) theory are tabulated~\cite{Minakata:2022yvs}. ``IA (X)'' and ``IB (Y)'' denote, respectively, shorthand of ``IA-SRP'' with $\theta_{12}$ non-flip and``IB-SRP'' with $\theta_{12}$ flip. 
}
\label{tab:SRP-symmetry}
\vglue 0.2cm
\begin{tabular}{c|c|c}
\hline 
SRP Symmetry & 
Vacuum parameter transformations & 
Matter parameter transformations
\\
\hline 
\hline 
IA (X) & 
none & 
$\lambda_{1} \leftrightarrow \lambda_{2}$, 
$c_{\varphi} \rightarrow \mp s_{\varphi}$, 
$s_{\varphi} \rightarrow \pm c_{\varphi}$. \\
\hline 
IB (Y) & 
$\theta_{12} \rightarrow - \theta_{12}$, 
$\delta \rightarrow \delta + \pi$. & 
$\lambda_{1} \leftrightarrow \lambda_{2}$, 
$c_{\varphi} \rightarrow \pm s_{\varphi}$, 
$s_{\varphi} \rightarrow \pm c_{\varphi}$. \\
\hline
IIA (Y) & 
$\theta_{23} \rightarrow - \theta_{23}$, 
$\theta_{12} \rightarrow - \theta_{12}$. & 
$\lambda_{1} \leftrightarrow \lambda_{2}$, 
$c_{\varphi} \rightarrow \pm s_{\varphi}$, 
$s_{\varphi} \rightarrow \pm c_{\varphi}$. \\
\hline 
IIB (X) & 
$\theta_{23} \rightarrow - \theta_{23}$, 
$\delta \rightarrow \delta + \pi$. & 
$\lambda_{1} \leftrightarrow \lambda_{2}$, 
$c_{\varphi} \rightarrow \mp s_{\varphi}$, 
$s_{\varphi} \rightarrow \pm c_{\varphi}$. \\
\hline 
IIIA (Y) & 
$\theta_{13} \rightarrow - \theta_{13}$, 
$\theta_{12} \rightarrow - \theta_{12}$. & 
$\lambda_{1} \leftrightarrow \lambda_{2}$, 
$c_{\varphi} \rightarrow \pm s_{\varphi}$, 
$s_{\varphi} \rightarrow \pm c_{\varphi}$. \\
\hline 
IIIB (X) & 
$\theta_{13} \rightarrow - \theta_{13}$, 
$\delta \rightarrow \delta + \pi$. & 
$\lambda_{1} \leftrightarrow \lambda_{2}$, 
$c_{\varphi} \rightarrow \mp s_{\varphi}$, 
$s_{\varphi} \rightarrow \pm c_{\varphi}$. \\
\hline 
IVA (X) & 
$\theta_{23} \rightarrow - \theta_{23}$, 
$\theta_{13} \rightarrow - \theta_{13}$ & 
$\lambda_{1} \leftrightarrow \lambda_{2}$, 
$c_{\varphi} \rightarrow \mp s_{\varphi}$, 
$s_{\varphi} \rightarrow \pm c_{\varphi}$. \\
\hline 
IVB (Y) & 
$\theta_{23} \rightarrow - \theta_{23}$, 
$\theta_{13} \rightarrow - \theta_{13}$, & 
$\lambda_{1} \leftrightarrow \lambda_{2}$, \\ 
 &
$\theta_{12} \rightarrow - \theta_{12}$, $\delta \rightarrow \delta + \pi$. 
 &
$c_{\varphi} \rightarrow \pm s_{\varphi}$, $s_{\varphi} \rightarrow \pm c_{\varphi}$. \\
\hline 
\end{tabular}
\end{center}
\vglue -0.6cm 
\end{table}

\begin{table}[h!]
\vglue 0.2cm
\begin{center}
\caption{All the symmetries ``Symmetry Z-DMP'' of the 1-2 state exchange type found in DMP are tabulated~\cite{Minakata:2021dqh}. The notations follow those in Table~\ref{tab:SRP-symmetry}.
}
\label{tab:DMP-symmetry}
\vglue 0.2cm
\begin{tabular}{c|c|c}
\hline 
DMP Symmetry & 
Vacuum parameter transformations & 
Matter parameter transformations
\\
\hline 
\hline 
IA (X) & 
none & 
$\lambda_{1} \leftrightarrow \lambda_{2}$, 
$c_{\psi} \rightarrow \mp s_{\psi}$, 
$s_{\psi} \rightarrow \pm c_{\psi}$. \\
\hline 
IB (Y) & 
$\theta_{12} \rightarrow - \theta_{12}$, 
$\delta \rightarrow \delta + \pi$. & 
$\lambda_{1} \leftrightarrow \lambda_{2}$, 
$c_{\psi} \rightarrow \pm s_{\psi}$, 
$s_{\psi} \rightarrow \pm c_{\psi}$. \\
\hline
IIA (Y) & 
$\theta_{23} \rightarrow - \theta_{23}$, 
$\theta_{12} \rightarrow - \theta_{12}$. & 
$\lambda_{1} \leftrightarrow \lambda_{2}$, 
$c_{\psi} \rightarrow \pm s_{\psi}$, 
$s_{\psi} \rightarrow \pm c_{\psi}$. \\
\hline 
IIB (X) & 
$\theta_{23} \rightarrow - \theta_{23}$, 
$\delta \rightarrow \delta + \pi$. & 
$\lambda_{1} \leftrightarrow \lambda_{2}$, 
$c_{\psi} \rightarrow \mp s_{\psi}$, 
$s_{\psi} \rightarrow \pm c_{\psi}$. \\
\hline 
IIIA (Y) & 
$\theta_{13} \rightarrow - \theta_{13}$, 
$\theta_{12} \rightarrow - \theta_{12}$. & 
$\lambda_{1} \leftrightarrow \lambda_{2}$, 
$\phi \rightarrow - \phi$, \\ 
 & & 
$c_{\psi} \rightarrow \pm s_{\psi}$, 
$s_{\psi} \rightarrow \pm c_{\psi}$ \\
\hline 
IIIB (X) & 
$\theta_{13} \rightarrow - \theta_{13}$, 
$\delta \rightarrow \delta + \pi$. & 
$\lambda_{1} \leftrightarrow \lambda_{2}$, 
$\phi \rightarrow - \phi$, \\ 
 & & 
$c_{\psi} \rightarrow \mp s_{\psi}$, 
$s_{\psi} \rightarrow \pm c_{\psi}$. \\
\hline 
IVA (X) & 
$\theta_{23} \rightarrow - \theta_{23}$, 
$\theta_{13} \rightarrow - \theta_{13}$. & 
$\lambda_{1} \leftrightarrow \lambda_{2}$, 
$\phi \rightarrow - \phi$, \\ 
 & & 
$c_{\psi} \rightarrow \mp s_{\psi}$, 
$s_{\psi} \rightarrow \pm c_{\psi}$. \\
\hline 
IVB (Y) & 
$\theta_{23} \rightarrow - \theta_{23}$, 
$\theta_{13} \rightarrow - \theta_{13}$, & 
$\lambda_{1} \leftrightarrow \lambda_{2}$, 
$\phi \rightarrow - \phi$, \\ 
 &
$\theta_{12} \rightarrow - \theta_{12}$, $\delta \rightarrow \delta + \pi$. 
 &
$c_{\psi} \rightarrow \pm s_{\psi}$, $s_{\psi} \rightarrow \pm c_{\psi}$. \\
\hline 
\end{tabular}
\end{center}
\vglue -0.4cm 
\end{table}

The readers must be puzzled about what happened. The presence or absence of $\theta_{12}$ in DMP and SRP second conditions, respectively, and the issue of $\varphi$ transformation to $\psi$ and vice versa do not appear to ``cancel'' with each other to produce the essentially identical SRP and DMP 1-2 exchange symmetries. However, let us pursue this possibility.  
As a first trial, we examine the case that the second conditions consist only of the matrix part in the both theories. We make the transformations $s_{\varphi} \rightarrow c_{\psi}, c_{\varphi} \rightarrow - s_{\psi}, a \rightarrow \Delta m^2_{ \text{ren} }$ in the ``matrix-SRP'', i.e., the pre-factor removed eq.~\eqref{2nd-condition-SRP}, to obtain the ``matrix-DMP'' second condition. The obtained solutions of the ``matrix-DMP'' second condition shows that the correct symmetry is reproduced for Symmetry X, X=IA, IIB, IIIB, and IVA. But, in the remaining Symmetry Y, Y=IB, IIA, IIIA, and IVB, we obtain the symmetry transformations with the upper and lower $\pm$ (or $\mp$) signs interchanged, indicating that an overall minus sign has to be introduced (by hand) to make Symmetry Y holds. Let us denote the above Symmetry X as ``X-type'' and Symmetry Y as ``Y-type''. 
Then, what we have observed is that the ``matrix-DMP'' second condition produced by the above $\varphi$ to $\psi$ transformations from the ``matrix-SRP''  successfully reproduces the symmetry of X-type: $c_{\psi} \rightarrow \mp s_{\psi}$ and $s_{\psi} \rightarrow \pm c_{\psi}$, but not of Y-type: $c_{\psi} \rightarrow \pm s_{\psi}$ and $s_{\psi} \rightarrow \pm c_{\psi}$. But, what is lacking is an overall minus sign. 

When we go back to the original second conditions with the pre-factors, this sign problem can be resolved if $\theta_{12}$ sign flip is involved in the Y-type symmetries and not in the X-type symmetries. We observe in Table~\ref{tab:SRP-symmetry} and Table~\ref{tab:DMP-symmetry} that it is indeed the case. This is the reason why the SRP and DMP symmetries are essentially identical. 

Then the question would be: Why does the $\theta_{12}$ sign flip occur only in the Y-type symmetries, namely, only when it is required? In DMP, it is just the solutions of the (full) second condition. But,  we are now trying to reproduce the DMP 1-2 symmetry from SRP to prove the identity of the two symmetries (or vice versa). Alas, in SRP neither the first nor the second conditions involve $\theta_{12}$, leaving no chance for triggering the $\theta_{12}$ sign flip from the SF equation. Then, why should $\theta_{12} \rightarrow - \theta_{12}$ be included in Table~\ref{tab:SRP-symmetry} for SRP in the Y-type symmetries? 

The answer is that without the $\theta_{12}$ transformation the consistency with the eigenvalue exchange $\lambda_{1} \leftrightarrow \lambda_{2}$ is lost in the Y-type symmetries. The zeroth-order eigenvalues in SRP are given by~\cite{Martinez-Soler:2019nhb,Minakata:2022yvs} 
\begin{eqnarray} 
\lambda_{1} 
&=& 
\sin^2 \left( \varphi - \theta_{12} \right) \Delta m^2_{21} + \cos^2 \varphi c^2_{13} a, 
\nonumber \\
\lambda_{2} 
&=&
\cos^2 \left( \varphi - \theta_{12} \right) \Delta m^2_{21} + \sin^2 \varphi c^2_{13} a, 
\nonumber \\
\lambda_{3} 
&=& 
\Delta m^2_{31} + s^2_{13} a. 
\label{eigenvalues-SRP}
\end{eqnarray}
Under the transformations $c_{\varphi} \rightarrow \pm s_{\varphi}$ and $s_{\varphi} \rightarrow \pm c_{\varphi}$ sine and cosine of $( \varphi - \theta_{12} )$ transform as 
\begin{eqnarray} 
&&
\sin \left( \varphi - \theta_{12} \right) 
\rightarrow 
\pm \cos ( \varphi + \theta_{12} ), 
\nonumber \\
&& 
\cos ( \varphi - \theta_{12} ) 
\rightarrow 
\pm \sin ( \varphi + \theta_{12} ). 
\nonumber 
\end{eqnarray}
Therefore, without including the sign flip of $\theta_{12}$, the transformations $c_{\varphi} \rightarrow \pm s_{\varphi}$ and $s_{\varphi} \rightarrow \pm c_{\varphi}$ are not consistent with $\lambda_{1} \leftrightarrow \lambda_{2}$. This problem does not occur in the X-type symmetries. It is the reason why the $\theta_{12}$ sign flip must be included in the Y-type symmetries in SRP. 
Thus, we have understood the reason why SRP and DMP theories possess the identical 1-2 exchange symmetries apart from difference in matter undressing and dressing of $\theta_{13}$. 
We note that the identity comes out via a highly nontrivial way, by overcoming the marked differences, in parts by parts, between the SF second conditions in the both theories. 

\section{Concluding remarks} 
\label{sec:conclusion} 

We have started by reviewing the current status of our understanding of the Rep (shorthand for ``reparametrization'') symmetry in neutrino oscillation in matter by using the SF (Symmetry Finder) method~\cite{Minakata:2022yvs,Minakata:2021dqh,Minakata:2021goi,Minakata:2022zua}. While non-negligible numbers of the Rep symmetries are identified to date, there exist still many unanswered questions. One of them is so called the ``problem of framework dependence'', which addresses the question of whether the Rep symmetry originates from the physics request, or characterized as a framework-dependent regularity as it may look more natural. The SF symmetry, by construction, looks like a framework dependent symmetry, but we have challenged to this common view by proposing the ``one-resonance - one-symmetry'' picture of the Rep symmetry. That is, the symmetry belongs to the resonance, a physical object, not the framework.

Can we confirm, or refute, the ``one-resonance - one-symmetry'' picture? It is known that the both SRP (solar-resonance perturbation) and DMP theories possess the 1-2 state exchange symmetries. SRP and DMP are, respectively, a locally-valid and a globally-valid theories (see section~\ref{sec:globally-valid}) whose regions of validity as well as the structures of perturbative formulation differ markedly. If we can show that these two 1-2 exchange symmetries are identical to each other, we can argue that the symmetry is associated with the resonance, not the framework.
In this paper we have undertaken such a comparative study of the symmetries, and indeed confirmed that the 1-2 exchange symmetries in SRP and DMP are identical to each other apart from matter-dressing or undressing of $\theta_{12}$.

Thus, we have learned by our comparative study of the symmetries that while Rep symmetry of the 1-2 state exchange type exists in the respective theories, SRP and DMP, with their own inherently different details, they both describe the identical object, the solar-scale matter enhanced oscillation. In this sense our ``one-resonance - one-symmetry'' picture passes the 1-2 symmetry comparative study test. 

\subsection{Problem of missing 1-3 symmetry in DMP} 
\label{sec:missing-1-3} 

We must remark, however, that the identity of the SRP and DMP 1-2 exchange symmetries poses us a serious question. One can argue that the same feature must prevail to the atmospheric resonance region. If it is the case, since the 1-3 exchange symmetry is known to exist in the locally-valid helio-perturbation theory~\cite{Minakata:2021goi}, it must exists also in the globally-valid DMP. 
To our current knowledge, DMP does have the 1-2 symmetry but {\em not} the 1-3 exchange symmetry, and we have the problem of missing 1-3 symmetry in DMP. Answering this intriguing question is left for future study. 

We note that, despite the obvious interests, the Rep symmetry has not been examined in the AKT perturbation theory~\cite{Agarwalla:2013tza}, the alternative globally-valid framework. Need for investigating symmetry in this theory becomes higher as we face the 1-3 symmetry problem, because the Hamiltonian is diagonalized first in the 1-2 and then the 1-3 rotations in the AKT formalism. It means that, as far as the matrix multiplication structure is concerned, DMP's 1-2 exchange symmetry corresponds to the AKT's 1-3 exchange symmetry. It suggests that AKT could be the better way to approach the 1-3 exchange symmetry. 

As described in ref.~\cite{Minakata:2022yvs}, our research field of the symmetry in neutrino oscillation is still in its infancy with only less than two years of SF search for the Rep symmetry. Naturally there exist many difficult and unanswered questions. They include: 
\begin{itemize}

\item 
How big is the Rep symmetry in a given theory? 
For a possible way of thinking, see ref.~\cite{Minakata:2022zua} for a bold conjecture. 

\item 
What is the reason why the symmetries exist? 
Does their existence universal in any reasonable theory of neutrino evolution?

\end{itemize}
We hope that we will be able to grasp hints for answering some of these questions in the near future. 

\begin{acknowledgments}

The author thanks Ivan Martinez-Soler for kindly providing an improved version of Fig.~\ref{fig:global-oscillation}. 

\end{acknowledgments}

\end{document}